\documentclass[referee,sn-mathphys,Numbered]{sn-jnl}

\usepackage{graphicx}%
\usepackage{multirow}%
\usepackage{amsmath,amssymb,amsfonts}%
\usepackage{amsthm}%
\usepackage{mathrsfs}%
\usepackage[title]{appendix}%
\usepackage{xcolor}%
\usepackage{textcomp}%
\usepackage{manyfoot}%
\usepackage{booktabs}%
\usepackage{algorithm}%
\usepackage{algorithmicx}%
\usepackage{algpseudocode}%
\usepackage{listings}%

\theoremstyle{thmstyleone}%
\theoremstyle{thmstyletwo}%

\theoremstyle{thmstylethree}%

\begin{document}

\title[New solution cosmological wormhole]{A new solution for a generalized cosmological wormhole}


\author*[1]{\fnm{Daniela} \sur{P\'erez}}\email{danielaperez@iar.unlp.edu.ar}

\author[2,3]{\fnm{M\'ario} \sur{Raia Neto}}\email{mrneto@estudante.ufscar.br}


\affil*[1]{\orgname{Instituto Argentino de Radioastronom\'ia (IAR, CONICET/CIC/UNLP)},  \city{Villa Elisa}, \postcode{C.C.5, (1894)}, \state{Buenos Aires}, \country{Argentina}}

\affil[2]{\orgdiv{Department of Physics}, \orgname{Federal University of São Carlos}, \city{S\~{a}o Carlos}, \postcode{13565-905}, \state{S\~{a}o Paulo}, \country{Brazil}}

\affil[3]{\orgname{Instituto Nacional de Pesquisas Espaciais}, \orgaddress{\street{Av. dos Astronautas, 1758, Jardim da Granja}, \city{São Jos\'e dos Campos}, \postcode{12227-010}, \country{Brazil}}}


\abstract{We find a new exact solution to Einstein field equations that represents a cosmological wormhole embedded in a flat Friedmann-Lemaître-Robertson-Walker universe. The new metric is a generalization of a previous cosmological wormhole solution found by Kim. We explicitly show that the flaring out condition is satisfied at the throat at all cosmic times; in addition, the null energy condition is violated at the throat regardless of the background cosmological model; thus, the spacetime geometry  presented here describes a wormhole coupled to the cosmic dynamics that exists at all cosmic times and whose throat remains open in any cosmological model.
}




\maketitle

\section{Introduction}

Wormholes are shortcuts between different regions of the universe. They are a special class of solutions of Einstein field equations, which represent multiply connected spacetimes. Wormholes require the absence of event horizons; the ``bridge'' connecting different events in spacetime has a minimum radius, the throat; in their simpler versions, two mouths on either each side of the throat allow the passage of matter and fields in both directions.

In the late eighties, wormhole research experienced a major revival after the publication of Morris and Thorne's solution that describes a traversable wormhole \cite{mor+88}. Since then, wormhole solutions have been the subject of intense theoretical study. Although their nature is still conjectural, many works have been devoted to the astrophysical consequences of their existence. An updated review of the literature on astrophysical wormholes can be found in \cite{bam+21}.

The Morris and Thorne (MT) wormhole solution is both static and asymptotically flat. Over the years, there have been many attempts to construct dynamic wormholes solutions and, in particular, wormholes that evolve due to their coupling with the cosmological background.

Roman \citep{rom93} investigated whether inflation might provide a natural mechanism to enlarge wormholes of microscopic size. For this purpose, he considered a Lorentzian wormhole embedded in a flat de Sitter space and showed that the throat expands at the same rate as the scale factor. At the same time, Kim \cite{kim92} considered the effects of the cosmological constant on a Lorentzian wormhole constructed by connecting two Schwarzschild-de Sitter spacetimes. Hochberg and Kepart \cite{hoc+93} also used a surgical procedure to obtain a wormhole solution from two copies of the Friedmann-Lemaître-Robertson-Walker (FLRW) metric (see also \citep{mae+09}). Another approach adopted to obtaining cosmological wormhole solutions is to add a time-dependent scaling factor to the metric \cite{kim96,kar+06,kuf16}.

Cosmological wormhole solutions have also been derived in alternative theories of gravity such as $f(R)$-gravity \cite{sai+11,bah+16}, $f(T)$-gravity \cite{sha+13} and Lovelock gravity \cite{meh+21}.One of these solutions \cite{bah+16} was recently used by Pavlović and Sossich \cite{pav+23} to analyze whether wormholes could be created during a cosmological bounce.


Among cosmological black hole spacetimes, the McVittie metric \cite{mcv33} is probably the one that has been most analyzed. It is an exact solution of Einstein field equations for a central inhomogeneity in a cosmological setting. Using a similar approach, Kim \citep{kim18,kim20} derived a new solution for a MT wormhole embedded in a FLRW background. As in McVittie's original proposal, the solution is constructed in such a way that accretion onto the wormhole is not allowed. In the present paper, we obtain a new metric for a cosmological wormhole that generalizes Kim's solution. We explicitly show that the new metric satisfies the criteria for a dynamical wormhole and that the energy conditions are always violated at the throat.

The paper is structured as follows: in section \ref{sec:2} we present the generalized cosmological wormhole metric and calculate Einstein field equations for the proposed ansatz. Next, in section \ref{sec:3}, we analyze a specific model for the generalized cosmological wormhole: we determine the location of the throat, compute the flare-out condition (\ref{sec:3a}) and Misner-Sharp-Hernandez mass (\ref{sec3:b}), and finally show that the null energy condition is violated at the throat (\ref{sec3:c}). We close the paper with some conclusions.

\section{Generalized cosmological wormhole metric}\label{sec:2}


The cosmological wormhole metric derived by Kim \cite{kim18} is characterized by a redshift factor equal to 1 ($\Phi' = 0$) and a shape function of the form $b(r) = b^2_0/r$, being $b_0$ the location of the wormhole throat. In isotropic coordinates, $\left(t,r,\theta,\phi\right)$, the line element reads \footnote{Throughout the paper, we use a geometric unit system, that is, $G = c = 1$.}
\begin{equation}\label{kim-metric}
ds^2 = -dt^2 + \frac{a^2(t)}{\left(1+k r^2\right)^2}\left(1+\frac{b^2_0}{4 r^2}\right)^2 \left(dr^2+r^2 d\Omega^2\right).
\end{equation}
Here, $a(t)$ is the scale factor, $k$ is the curvature of space, $t$ is the cosmic time, and $d\Omega^2 \equiv d\theta^{2} + \sin^{2}{\theta} d\phi^{2}$ is the line element of the unit two-sphere. The coordinate $r$ is defined in the range $ 0 < r < \infty $. In what follows, we focus on flat cosmological models ($k = 0$). 

We see that if $a^2(t) \equiv 1$, we recover the metric for a Morris-Thorne wormhole \citep{mor+88},
\begin{equation}\label{mt}
ds^2 = -dt^2 + \left(1+\frac{b^2_0}{4 r^2}\right)^2 \left(dr^2+r^2 d\Omega^2\right),
\end{equation}
while by setting $b_0 = 0$ we obtain the FLRW line element
\begin{equation}
ds^2 = -dt^2 + a^2(t) \left(dr^2+r^2 d\Omega^2\right).
\end{equation}

We make the following proposal for the generalized cosmological wormhole metric
\begin{equation}\label{gcw}
ds^{2} = -dt^2 + a^2(t)\left(1+\frac{b(r,t)^2}{4 r}\right)^2 \left(dr^2+r^2 d\Omega^2\right), \end{equation}
where $b(r,t) = b_1(r) \times b_2(t)$, being $b_2(t)$ an unspecified function of the cosmic time. In particular, we choose for the radial part of the shape function
\begin{equation}
b_1(r) = \frac{b_0}{\sqrt{r}}.  
\end{equation}

Notice that we recover the line element of 
\begin{itemize}
\item  the MT wormhole  taking $a(t) = 1$, $b_{1}(r) = b_0/\sqrt{r}$ and $b_{2}(t) = 1$.
\item Kim's cosmological wormhole choosing $b_{1}(r) = b_0/\sqrt{r}$ and $b_{2}(t) = 1$.
\end{itemize}
We next compute the components of the Einstein tensor ${G^{\mu}}_{\nu}$ for the spacetime metric \eqref{gcw}.

\subsection{Einstein field equations}

The mixed components of the Einstein tensor ${G^{0}}_{0}$, ${G^{1}}_{1}$, ${G^{2}}_{2}$ and ${G^{3}}_{3}$ can be divided into 3 different parts, each with a specific physical interpretation:
\begin{equation}
{G^{\mu}}_{\nu} = \left.{G^{\mu}}_{\nu}\right\vert_{\mathrm{c}} + {G^{\mu}}_{\nu}{\rvert}_{\mathrm{kcw}} + {G^{\mu}}_{\nu}{\rvert}_{\mathrm{gcm}}.
\end{equation}
Here, ``c'', ``kcw'' and ``gcw'' are the acronyms for cosmological, Kim cosmological wormhole and generalized cosmological wormhole, respectively. The explicit expressions are
\begin{eqnarray}
{G^{0}}_{0} &=& \left. - 3\frac{a'(t)}{a(t)}\right\vert_{\mathrm{c}} +\left. \frac{256 \: b^2_0 \; r^4 \; b^2_{2}(t)}{a^2(t) \alpha(r,t)^4}\right\vert_{\mathrm{kcw}}
 -  \left. 12  \frac{ b_0 \; b_2(t)\;  b'_{2}(t)\;  f_{00}(r,t)}{a(t) \alpha(r,t)^2}\right\vert_{\mathrm{gcw}},\\
{G^{1}}_{1} & = & \left.\left[- \frac{a'^2(t)}{a^2(t)} - 2 \frac{a''(t)}{a(t)}\right]\right\vert_{\mathrm{c}}
 -  \left. \frac{256 \: b^2_0 \; r^4 \; b^2_{2}(t)}{a^2(t) \alpha(r,t)^4}\right\vert_{\mathrm{kcw}}
-  \left.4 \left[\frac{b'_2(t) f_{11}(t,t)}{a(t) \; \alpha(r,t)^2} + \frac{b^2_0 \; b_2(t) \;  b''_2(t)}{\alpha(r,t)}\right]\right\vert_{\mathrm{gcw}}.\\
{G^{2}}_{2} & = & {G^{3}}_{3} = \left.\left[- \frac{a'^2(t)}{a^2(t)} - 2 \frac{a''(t)}{a(t)}\right]\right\vert_{\mathrm{c}}
 +  \left. \frac{256 \: b^2_0 \; r^4 \; b^2_{2}(t)}{a^2(t) \alpha(r,t)^4}\right\vert_{\mathrm{kcw}}
 -  \left.4 \left[\frac{b'_2(t) f_{11}(t,t)}{a(t) \; \alpha(r,t)^2} + \frac{b^2_0 \; b_2(t) \;  b''_2(t)}{\alpha(r,t)}\right]\right\vert_{\mathrm{gcw}},
\end{eqnarray}
where
\begin{eqnarray}
f_{00}(r,t) & = & \alpha(r,t) a'(t)+ b^2_0(t) \; a(t) \; b_2(t)\;  b'_{2}(t),\\
f_{11}(r,t) & = & \; 3 b_2(t) \; \alpha(r,t) \; a'(t) 
 +  2 \; a(t) \left[2 r^2 +b^2_0(t) b^2_2(t)\right] b'_2(t),\\
\alpha(r,t) & = & 4 r^2 + b^2_0 \; b^2_{2}(t).
\end{eqnarray}

The presence of the factor $b_2(t)$ in the metric gives the Einstein tensor an additional nonzero component 
\begin{eqnarray}
{G^{0}}_{1} & = & -32 \frac{b^2_0 \; r \; b_2(t) \; b'_2(t)}{\alpha^2(r,t)},\\
{G^{1}}_{0} & = & 64  \frac{b^2_{0} \; r^5 \; b_2(t) \; b'_2(t)}{a^2(t) \alpha^4(r,t)}.
\end{eqnarray}

Next, we consider an imperfect fluid with energy-momentum tensor given by
\begin{equation}\label{emt}
T_{ab} = \left(\rho +p\right) u_{a}u_{b} + p g_{ab} + q_{a}u_{b}+q_{b}u_{a},
\end{equation}
where $\rho$ is the density, $p$ is the pressure, $u^a$ is the four-velocity of the fluid, and $q^a$ is a spatial vector field that represents the current density of heat. We further assume that
\begin{equation}
u^{\mu} = \left(1,0,0,0\right),\;\;\; q^{\alpha}= \left(0,q,0,0\right), \;\;\;u^{b}q_{b} = 0. 
\end{equation}
Both density and pressure have two distinct components: one corresponding to the cosmological fluid and another associated with the matter that constitutes the wormhole, that is
\begin{eqnarray}
\rho(r,t) & = & \rho_{\mathrm{c}}(t) + \rho_{\mathrm{gcw}}(r,t),\\
p(r,t) & = & p_{\mathrm{c}}(t) + p_{\mathrm{gcw}}(r,t).
\end{eqnarray}

Einstein field equations take the form
\begin{multline}
\left.  -  3\frac{a'^2(t)}{a^2(t)}\right\vert_{\mathrm{c}} +\left. \frac{256 \: b^2_0 \; r^4 \; b^2_{2}(t)}{a^2(t) \alpha(r,t)^4}\right\vert_{\mathrm{kcw}}
 -  \left.12 \frac{b^2_0 \; b_2(t)\;  b'_{2}(t)\;  f_{00}(r,t)}{a(t) \alpha(r,t)^2}\right\vert_{\mathrm{gcw}} \\ 
 =  8 \pi \left(- \rho_{\mathrm{c}}(t) - \rho_{\mathrm{gcw}}(r,t)\right),\label{g00}
\end{multline}
\begin{multline}
\left.\left[- \frac{a'^2(t)}{a^2(t)} - 2 \frac{a''(t)}{a(t)}\right]\right\vert_{\mathrm{c}}
 -  \left. \frac{256 \: b^2_0 \; r^4 \; b^2_{2}(t)}{a^2(t) \alpha(r,t)^4}\right\vert_{\mathrm{kcw}}
 - \left. 4 \left[\frac{b'_2(t) f_{11}(t,t)}{a(t) \; \alpha(r,t)^2} + \frac{b^2_0 \; b_2(t) \;  b''_2(t)}{\alpha(r,t)}\right]\right\vert_{\mathrm{gcw}}\\
 =  8 \pi 
\left(p_{\mathrm{c}}(t)  +  pr_{\mathrm{gcw}}(r,t) \right),\label{g11}
\end{multline}
\begin{multline}
\left.\left[- \frac{a'^2(t)}{a^2(t)} - 2 \frac{a''(t)}{a(t)}\right]\right\vert_{\mathrm{c}}
 +  \left. \frac{256 \: b^2_0 \; r^4 \; b^2_{2}(t)}{a^2(t) \alpha(r,t)^4}\right\vert_{\mathrm{kcw}}
 - \left. 4 \left[\frac{b'_2(t) f_{11}(t,t)}{a(t) \; \alpha(r,t)^2} + \frac{b^2_0 \; b_2(t) \;  b''_2(t)}{\alpha(r,t)}\right]\right\vert_{\mathrm{gcw}}\\
 = 8 \pi 
\left(p_{\mathrm{c}}(t)  +  pt_{\mathrm{gcw}}(r,t) \right),\label{g22}
\end{multline}
\begin{equation}
- 64 \frac{b^2_0 \; r^5 \; b_2(t) b'_2(t)}{ \pi \; a^2(t) \alpha^4(r,t)}  =   q \label{g01}.
\end{equation}


Equations \eqref{g00}-\eqref{g22} show  that the wormhole and the cosmological background can be decoupled. We recover the Friedmann-Robertson equations by setting $b_0 = 0$. In the case $b_{2} = 1$, we get Kim cosmological wormhole metric and $q = 0$ (the energy-momentum tensor is that of a perfect fluid). We see that the addition of the factor $b_2(t)$ in \eqref{gcw} allows a current density of heat $q$ through the wormhole.  

In the following, we analyze the properties of the metric \eqref{gcw} for a specific choice of the function $b_2(t)$.


\section{Specific model for the generalized cosmological wormhole}\label{sec:3}

We now assume $b_2(t) \equiv 1/a(t)$. The line element in isotropic coordinates takes the form
\begin{equation}\label{23}
ds^{2} = -dt^2 + a^2(t)\left(1+\frac{b^2_0}{4 a^{2}(t)r^2}\right)^2 \left(dr^2+r^2 d\Omega^2\right).   
\end{equation}
The choice of the redshift function ($\Phi'=0$) ensures the absence of event horizons. 

We identify the throat following the definition introduced by Hochberg and Visser \citep{hoc+98b} for the case of dynamical wormholes. According to these authors, we can locate the throat by looking for a certain behaviour of the null geodesics of the metric: the throat is a minimal two-surface where null rays coming from the mouth focus and, once they pass the throat, they start to expand to the other side. If we denote by $n^{a}$ and $l^{a}$ the ingoing and outgoing tangent fields of null radial geodesics, and by $\theta_{n}$ and $\theta_{l}$ their respective expansions, one of the following conditions is satisfied on the throat \citep{mcn+21}
\begin{equation}\label{condw1}
\theta_{n} = 0 \;\; \wedge \;\; n^{a}\nabla_{a}\theta_{n} \ge 0,
\end{equation}
or,
\begin{equation}\label{condw2}
\theta_{l} = 0 \;\; \wedge \;\; l^{a}\nabla_{a}\theta_{l} \ge 0.
\end{equation}
The condition $n^{a}\nabla_{a}\theta_{n} \ge 0$ ($l^{a}\nabla_{a}\theta_{l} \ge 0$) is the generalization of the Morris-Thorne flare-out condition for dynamical wormholes.

Next, we compute whether conditions \eqref{condw1} or \eqref{condw2} is satisfied. But first, we will perform a coordinate transformation that it is better suited for our purpose.

\subsection{Throat and flare-out condition}\label{sec:3a}

We write the line element in terms of the areal radius coordinate $R$
\begin{equation}
R  =  a(t) \left(1+ \frac{b^2_0}{4 a^2(t) r^2}\right) r.\label{areal_radius}
\end{equation}
as
\begin{equation}\label{dt_dR}
ds^2  =  -\left(1 - R^2 H^2\right) dt^2 + \frac{1}{\left(1 - \frac{b^2_0}{R^2}\right)} dR^2  -  2 \frac{R H}{\sqrt{1-\frac{b^2_0}{R^2}}} dR dt+  R^2 d\Omega^2,    
\end{equation}
where $H \equiv a'(t)/a(t)$ is the Hubble factor.

Because of the spherical symmetry, the equation for the ingoing and outgoing radial null geodesics can be derived by setting $d\theta = d\phi = 0$ in $ds^{2}= 0$, thus obtaining
\begin{equation}
\left. \frac{dR}{dt}\right\vert_{\pm} = \left(\pm 1 + R H\right) \sqrt{1- \frac{b^2_0}{R^2}},
\end{equation}
where the ``$ - $'' (``$+$'') corresponds to ingoing (outgoing) case. The tangent vector fields $n^{a}$ and $l^{a}$ have the form
\begin{eqnarray}
n^{\mu} & = & \left(1,\left(- 1 + R H\right) \sqrt{1- \frac{b^2_0}{R^2}} ,0,0\right),\label{na}\\
l^{\mu} & = & \left(1,\left(+ 1 + R H\right) \sqrt{1- \frac{b^2_0}{R^2}},0,0\right).\label{la}
\end{eqnarray}
The expansion of the null vector $n^{a}$ when the geodesic to which it is tangent is not necessarily affinely-parametrized can be computed using the expression \citep{far15}
\begin{equation}
\theta_{n} = \left[ g^{ab} + \frac{l^{a} n^{b} + n^{a} l^{b}}{- n^{c} l^{d} g_{cd}}\right] \nabla_{a} n_{b}.
\end{equation}
In the same way, the expansion of the null vector $l^{a}$ obeys the relation
\begin{equation}
\theta_{l} = \left[ g^{ab} + \frac{l^{a} n^{b} + n^{a} l^{b}}{- n^{c} l^{d} g_{cd}}\right] \nabla_{a} l_{b}.
\end{equation}

Given these definitions and under the choice of $n^{a}$ and $l^{a}$ (Eqs. \eqref{na} and \eqref{la}), condition \eqref{condw1} yields
\begin{equation}
\theta_{n} =\frac{2 \sqrt{1-\frac{b^2_0}{R}} \left(-1 + R H\right)}{R} = 0, \; \Rightarrow \; R = \pm b_0, \;\; R = 1/H. 
\end{equation}
Thus, $R = b_0$ corresponds to the throat and $R = 1/H$ to the cosmological horizon of a spatially flat FLRW space. We see that the flaring-out condition is satisfied on the throat
\begin{equation}
\left. n^{a}\nabla_{a}\theta_{n} \right\vert_{R = b_0} , \;\; \Rightarrow \;\; 2 \frac{\left(-1+b_0 H\right)^{2}}{b^2_0} \ge 0.
\end{equation}
Condition \eqref{condw2} gives a similar result
\begin{equation}
\theta_{l} =\frac{2 \sqrt{1-\frac{b^2_0}{R}} \left(1 + R H\right)}{R} = 0, \; \Rightarrow \; R = \pm b_0, 
\end{equation}
and the flaring-out condition is also satisfied on the throat
\begin{equation}
\left. l^{a}\nabla_{a}\theta_{l} \right\vert_{R = b_0} , \;\; \Rightarrow \;\; 2 \frac{\left(1+b_0 H\right)^{2}}{b^2_0} \ge 0.
\end{equation}

Thus, the spacetime geometry given by the line element \eqref{23} represents a cosmological wormhole with a throat at $R = b_0$ and a cosmological horizon at $R = 1/H$.

We also calculate the location of the throat in terms of the isotropic radius $r$ by solving the equation
\begin{equation}
b_0  =  a(t) \left(1+ \frac{b^2_0}{4 a^2(t) r^2}\right) r. \end{equation}
The solution is $r = b_0/\left(2 a(t)\right)$. We see that regardless of the coordinates we use ($R$ or $r$), there is always a throat for any background cosmological model. This is not the case in the Kim cosmological wormhole solution \citep{kim18,kim20}: the existence of the throat depends on the peak value of the energy density of the wormhole. 


\subsection{Misner-Sharp-Hernandez mass}\label{sec3:b}

The Misner-Sharp-Hernandez (MSH) mass is a quantity that allows to identify localized sources of gravity \citep{mis+64,her+66}; it is defined only in spherically symmetric spacetimes and, in this case, coincides with the Hawking-Hayward quasi-local energy \citep{haw68}. We can write the line element in terms of the areal radius coordinate $R$ and the angular coordinates $(\theta, \phi$) as follows \citep{far15}
\begin{equation}
ds^2 = h_{ab} dx^a \; dx^b + R^2 d\Omega^2,
\end{equation}
where $x^a = (t,R)$ and 
\begin{equation}
h_{ab}  =  \mathrm{diag}\left[-1, a^2(t) \left(1+ \frac{b^2_0}{4 a^2(t) r^2}\right)\right].
\end{equation}
The Misner-Sharp-Hernandez mass $M_{\mathrm{MSH}}$ is defined as
\begin{equation}
M_{\mathrm{MSH}} = \frac{R}{2}\left(1 - h^{ab} \nabla_{a}R \nabla_{b} R\right).
\end{equation}
In our case, it takes the form\footnote{We use the identity
\begin{equation}
1 - \frac{b^2_0}{R^2} = \frac{\left(1-\frac{b^2_0}{4 a^{2}(t)\tilde{r}}^2\right)^2}{\left(1+\frac{b^2_0}{4 a^{2}(t)\tilde{r}}^2\right)^2}
\end{equation}
to reduce some of expressions obtained.}
\begin{equation}
M_{\mathrm{MSH}} = \frac{R}{2}\left[H^2 \left(R^2 -b^2_0\right) + \frac{b^2_0}{R^2}\right].
\end{equation}
The MSH mass can be computed in terms of the Riemann tensor. The decomposition of the latter into a Ricci and a Weyl part results into a natural decomposition of the MSH mass into a Ricci and a Weyl part \citep{car+10a, car+10b}. The Ricci part of the $M_{\mathrm{MSH}}$ is
\begin{equation}
E_{\mathrm{R}} = \frac{R^3 \; H^2}{2} \left(1 - \frac{b^2_0}{R^2}\right),
\end{equation}
and the corresponding Weyl part is
\begin{equation}
E_{\mathrm{W}} = \frac{b^2_0}{2R}.
\end{equation}
In the limit $b_0 \rightarrow 0$, we recover the expressions of the $M_{\mathrm{MSH}}$ in FLRW spacetime, that is, $M_{\mathrm{MSH}} = E_{\mathrm{R}} = R^3 \; H^2 /2$. If $a(t) \rightarrow 1$, $M_{\mathrm{MSH}} = E_{\mathrm{W}} = b^2_0 / 2R$, as in a MT wormhole.


\subsection{Energy conditions}\label{sec3:c}

Hochberg and Visser  showed that if the flaring-out condition is fulfilled at the throat, then the stress-energy tensor at the throat must  satisfy \citep{hoc+98b}
\begin{equation} \label{NEC}
T_{ab} k^{a} k^{b} \le 0,
\end{equation}
where $k^{a}$ is any null vector. Hence, the null energy condition (NEC) is violated at the throat. These authors also argued that not only static wormholes but also dynamic wormholes have NEC violations at the throat \citep{hoc+98a}. 

Following the method explained in \citep{poi04}, we will now derive the explicit expressions for the NEC in the case of the imperfect fluid given by \eqref{emt}. The corresponding energy-momentum tensor admits the decomposition
\begin{equation}
T^{\alpha \beta}  =  \rho \; \hat{e}^{\alpha}_{0} \hat{e}^{\beta}_{0} + p_1 \; \hat{e}^{\alpha}_{1}  \hat{e}^{\beta}_{1} +  p_2 \; \hat{e}^{\alpha}_{2}  \hat{e}^{\beta}_{2} + p_2 \; \hat{e}^{\alpha}_{3}  \hat{e}^{\beta}_{3} \nonumber
 +  q \; \hat{e}^{\alpha}_{0}  \hat{e}^{\beta}_{1} + q \; \hat{e}^{\beta}_{0} \hat{e}^{\alpha}_{1},      
\end{equation}
where the vectors $\hat{e}^{\alpha}_{\mu}$ form an orthonormal basis and obey the relations
\begin{equation}
g_{\alpha \beta} \hat{e}^{\alpha}_{\mu} \hat{e}^{\beta}_{\nu} = \eta_{\mu \nu}.
\end{equation}
Here, $\eta_{\mu \nu} = \mathrm{diag}\left(-1,1,1,1\right)$ is the Minkowski metric. We write the future-directed null vector $k^{\alpha}$ as
\begin{equation}
k^{\alpha} = \hat{e}^{\alpha}_{0}  + a \; \hat{e}^{\alpha}_{1}  + b\; \hat{e}^{\alpha}_{2} + c \; \hat{e}^{\alpha}_{3},  
\end{equation}
where $a$, $b$ and $c$ are arbitrary functions of the coordinates such that $a^2 + b^2 + c^2 = 1$. After some algebraic operations, the NEC yields
\begin{equation}
T_{ab} k^{a} k^{b} = \rho - 2 a q + a^2 p_1 + p_2 (b^2+c^2) \ge 0.
\end{equation}
If we choose $b = c = 0$, then $a = 1$ and we get
\begin{equation}
\rho + p_1 - 2 q \ge 0.
\end{equation}
Instead, if $a = c = 0$ and $b = 1$, the NEC reads
\begin{equation}
\rho + p_2 \ge 0.
\end{equation}
We get the same result for $c = 1$ and $a = b = 0$.

In short, the NEC in terms of the density $\rho$, principal pressures $p_1$ and $p_2$ and the heat flux $q$ is satisfied if
\begin{equation}\label{NEC}
\rho + p_1 - 2 q \ge 0, \; \; \; \rho + p_2 \ge 0.
\end{equation}

From Eqs. \eqref{g00}-\eqref{g01}, we get expressions for $\rho_{\mathrm{c}}$, $\rho_{\mathrm{gcw}}$, $pr_{\mathrm{c}}$, $pr_{\mathrm{gcw}}$,  $pt_{\mathrm{gcw}}$ and $q$:
\begin{eqnarray}
\rho_{\mathrm{c}} & = & \frac{3}{8 \pi}\frac{a'^2(t)}{a^2(t)},\\
\rho_{\mathrm{gcw}} & = - & \frac{2 b^2_0 r^2 \tilde{\rho}(r,t)}{\pi \beta^4},\\
p_{\mathrm{c}} & = & \frac{1}{8\pi}\left(- \frac{a'^2(t)}{a^2(t)}- 2 \frac{a''(t)}{a(t)}\right),\\
pr_{\mathrm{gcw}} & = & \frac{b^2_0  \tilde{pr}(r,t)}{2 \pi a^2(t) \beta^4},\\
pt_{\mathrm{gcw}} & = & \frac{b^2_0  \tilde{pt}(r,t)}{2 \pi a^2(t) \beta^4},\\
q & = & \frac{64 b^2_0 \; r^5 a'(t) \; a^3(t)}{\pi \beta^4},
\end{eqnarray}
where
\begin{eqnarray}
\tilde{\rho}(r,t) & = & 16 r^2 a^4(t) + 3 \beta^2 a'(t)^2,\\
\tilde{pr}(r,t) & = & -64 r^4 a^6(t) + \beta^2  \chi,\\
\tilde{pt}(r,t) & = &64 r^4 a^6(t) +\beta^2  \chi,\\
\beta & = & 4 a^2 (t) r^2 + b^2_0,\\
\chi & = & - b^2_0 a'^2(t) + a(t) \beta a''(t).
\end{eqnarray}

We now calculate the left-hand side of the inequalities \eqref{NEC} and evaluate the corresponding expressions at the throat ($r_{\mathrm{th}} = b_0 / (2 a(t))$
\begin{equation}
\rho + p_1 - 2 q  =  \left. \rho_{\mathrm{c}}+ \rho_{\mathrm{gcw}} + p_{\mathrm{c}} + pr_{\mathrm{gcw}} - 2 q \right\vert_{r_{\mathrm{th}}} 
 =  - \frac{a^2(t)+ b_0 a'(t) \left(1 + b_0 a'(t)\right)}{4 b^2_0 \pi a^2(t)} < 0,\label{necf1}
\end{equation}
if $a'(t) > 0$, which is the case in an expanding universe. The other null energy condition is 
\begin{equation}
\rho + p_2  =  \left. \rho_{\mathrm{c}}+ \rho_{\mathrm{gcw}} + pt_{\mathrm{gcw}} \right\vert_{r_{\mathrm{th}}} 
=   - \frac{a'^2(t)}{4 \pi a^2(t)} < 0.\label{necf2}
\end{equation}
The latter inequality shows that the NEC is violated independently of the background cosmological model.

Note that there are wormhole solutions where the energy conditions are satisfied. For example, it has been shown \cite{dai+18} that a positive cosmological constant can provide a static wormhole configuration. Dai and collaborators \cite{dai+20} also proved that a wormhole-like structure can form where the brane tension provides the repulsion that counteracts the gravitational attraction. In both cases, the wormhole solutions do not require exotic matter.

\section{Conclusions}

We have found a new solution to Einstein field equations that represents a MT wormhole embedded in a flat FLRW cosmological background. The obtained metric is a generalization of Kim's cosmological wormhole: the shape function depends on the radial coordinate and also on the cosmic time; we have considered that the source of the spacetime geometry is an imperfect fluid, so there is a current heat density.

We have also analyzed a specific model of the generalized cosmological wormhole solution. For this case, we have explicitly shown that the flaring out condition is always satisfied at the throat; in addition, we have demonstrated that the null energy condition is violated at the throat for all cosmic times, and for any scale factor.

An important difference between Kim's cosmological wormhole and the generalization presented here is that in our case the existence of the throat, and hence of the wormhole, is independent of the background cosmological model; this is not the case in Kim's solution where the throat of the wormhole comes into existence at a given cosmic time, or disappears as the universe evolves, depending on the background scale factor chosen. 

As stated earlier, wormholes are conjectural objects. However, since they are solutions to the Einstein field equations, they enable us to push the limits of General Relativity and explore the underlying principles of the theory. In a cosmological context, wormholes coupled to the cosmic dynamics could give rise to remarkable phenomena; for instance, intra-cosmological wormholes could bring two separate regions of the universe, at different cosmic epochs, into causal contact. In this sense, intra-cosmological wormholes could serve as bridges not only across space but also across cosmological eras. It is impossible to foresee all the implications of any theory, let alone such a vast and rich theory like General Relativity. However, we think that some of these issues deserve to be explored, and so we hope to do in the near future.

\backmatter

\bmhead{Acknowledgments}
D. P. is very grateful to Gustavo E. Romero and Santiago E. Perez Bergliaffa for the many insightful comments on this article. D. P. acknowledges the support from CONICET under Grant No. PIP 0554 and AGENCIA I$+$D+$i$ under Grant PICT-2021-I-INVI-00387. M.R.N. would like to thank his advisor and friend, Prof. Dr. Luiz Claudio Lima Botti, for the support on scientific research and friendship; the Instituto Nacional de Pesquisas Espaciais (INPE), Universidade Federal de São Carlos (UFSCar) and Centro de Radioastronomia e Astrofísica Mackenzie (CRAAM). M.R.N. is very grateful to the support of CNPq-PIBIC-INPE process: $133024/2021-0$

\section*{Declarations}


\begin{itemize}
\item Funding

D. P. acknowledges the support from CONICET under Grant No. PIP 0554 and AGENCIA I$+$D+$i$ under Grant PICT-2021-I-INVI-00387. M.R.N. acknowledges the support of CNPq-PIBIC-INPE process: $133024/2021-0$ 

\item Competing interests 

The authors have no competing interests to declare that are relevant to the content of this article.

\item Ethics approval 

Not applicable

\item Consent to participate

Not applicable

\item Consent for publication

Not applicable

\item Availability of data and materials

Not applicable

\item Code availability 

Not applicable

\item Authors' contributions

All authors contributed equally to the manuscript.

\end{itemize}







\bibliography{sn-bibliography}

\end{document}